\documentclass[journal=jctcce,manuscript=letter]{achemso}

\usepackage{graphicx}
\usepackage{bbm}
\usepackage{amsmath,amssymb}

\newcommand{\be}{\begin{equation}}
\newcommand{\beq}{\begin{equation}}
\newcommand{\ee}{\end{equation}}
\newcommand{\bea}{\begin{eqnarray}}
\newcommand{\eea}{\end{eqnarray}}
\newcommand{\ba}{\begin{array}}
\newcommand{\ea}{\end{array}}

\newcommand{\UAM}{Departamento de F\'{i}sica Te\'{o}rica de la Materia Condensada, Universidad Aut\'{o}noma
                  de Madrid  Campus de Cantoblanco Madrid 28049, Spain}
\newcommand{\ICMM}{Instituto de Ciencia de Materiales de Madrid, CSIC 28049 Madrid, Spain}
\newcommand{\TAMP}{Department of Physics, Tampere University of Technology, FI-33101 Tampere, Finland}
\newcommand{\UCIRVINE}{Department of Chemistry, University of California, Irvine, California 92697, USA}
\newcommand{\CNRMODENA}{CNR-Istituto di Nanoscienze Via Campi 213A, I-41125 Modena, Italy}
\newcommand{\ilm}{Institut Lumi\`ere Mati\`ere, UMR5306 Universit\'e Lyon 1-CNRS, Universit\'e de Lyon,
                   F-69622 Villeurbanne Cedex, France}
\author{J. G. Vilhena}
\affiliation{\ICMM}
\altaffiliation{\UAM}
\email{guilherme.vilhena@uam.es}
\author{E. R{\"a}s{\"a}nen}
\email{esa.rasanen@tut.fi}
\affiliation{\TAMP}
\author{M.\,A.\,L. Marques}
\affiliation{\ilm}
\author{S. Pittalis}
\affiliation{\CNRMODENA}
\altaffiliation{\UCIRVINE}
\title{Construction of the B88 exchange-energy functional in two dimensions}
\begin{document}

%%%%%%%%%%%%%%%%%%%%%%%%%%%%%%%%%%%%%%%%%%%%%%%%%%%%%%%%%%%%%%%%%%%%%%%
% Abstract
%%%%%%%%%%%%%%%%%%%%%%%%%%%%%%%%%%%%%%%%%%%%%%%%%%%%%%%%%%%%%%%%%%%%%%%
\begin{abstract}
We construct a generalized-gradient approximation for the 
exchange-energy density of finite two-dimensional systems. Guided by 
non-empirical principles, we include the proper small-gradient limit
and the proper tail for the exchange-hole potential. The observed 
performance is superior to that of the two-dimensional local-density
approximation, which underlines the usefulness of the approach in 
practical applications.
\end{abstract}
%%%%%%%%%%%%%%%%%%%%%%%%%%%%%%%%%%%%%%%%%%%%%%%%%%%%%%%%%%%%%%%%%%%%%%%

%\pacs{71.15.Mb, 31.15.E-, 73.21.La}
 \maketitle
 
%%%%%%%%%%%%%%%%%%%%%%%%%%%%%%%%%%%%%%%%%%%%%%%%%%%%%%%%%%%%%%%%%%%%%%%
%%%%%%%%%%%%%%%%%%%%%%%%%%%%%%%%%%%%%%%%%%%%%%%%%%%%%%%%%%%%%%%%%%%%%%%
% Introduction
%%%%%%%%%%%%%%%%%%%%%%%%%%%%%%%%%%%%%%%%%%%%%%%%%%%%%%%%%%%%%%%%%%%%%%%
%%%%%%%%%%%%%%%%%%%%%%%%%%%%%%%%%%%%%%%%%%%%%%%%%%%%%%%%%%%%%%%%%%%%%%%
\section{Introduction}
Nanoscale electronic devices define a large variety of low-dimensional
systems that range from atomistic to artificial structures. These 
include, e.g., modulated semiconductor layers and surfaces, quantum 
Hall systems, spintronic devices, quantum dots~\cite{qd} (QDs), 
quantum rings, and artificial graphene.~\cite{agreview} The complex
effects of electron-electron interactions pose a challenge to 
accurately compute the energy components of these structures.

Density-functional theory~\cite{dft1,dft2,dft3} (DFT) is ideally suited 
to balance numerical effort and accuracy. Considerable advances beyond 
the commonly used local-density approximation (LDA) have been achieved 
by generalized gradient approximations (GGAs), orbital functionals, and
hybrid functionals~\cite{functionals}. Previous studies have shown that 
most functionals developed for 3D systems break down when applied to 
realistic models of two-dimensional (2D) systems.~\cite{kim,pollack} 
In particular, accurate modeling of semiconductor quantum dots (in, e.g.
GaAs/AlGaAs interfaces~\cite{qd}) requires the use of 2D functionals, since
the degrees of freedom are suppressed in the direction perpendicular
to the plane. The relevance 
of including in standard 3D functionals the ability to recover the 2D 
limit -- at least at the LDA level -- has been clearly  demonstrated in
a recent work dealing with heterogeneous 3D atomistic materials.~\cite{chiodo}
The construction of more elaborated approximations for the 
exchange-correlation energy in 2D beyond LDA~\cite{rajagopal,tanatar,attaccalite} 
started also relatively recently~\cite{2d1,2d4,2d5,2d6,2d7,2d12}; in 
particular, they demonstrated some of the limitations of the 2D-LDA and
how to overcome them.

In this work we focus on exchange energies of finite systems and take 
the natural step beyond LDA by including the dependence of the 
functional on density gradients. We follow a procedure which solves the
long-standing challenge of obtaining an {\em non-empirical} gradient 
expansion for the exchange energy of finite 2D systems.~\cite{kirzhnits} 
We achieve  this result by carrying out a semiclassical analysis 
analogous to that of high-Z atoms in 3D.~\cite{b88,elliott} The form of 
the functional used as a paradigm is B88.~\cite{b88} This allows us not 
only to come up with a form that has a proper small-gradient expansion,
but also to obtain a model for the exchange-hole potential that has the
proper asymptotic tail.

The present work is organized as follows. In Sec.~\ref{theory} we briefly 
review the construction of the B88 functional (in 3D) and then proceed 
with the 2D case, exploiting the semiclassical limit of parabolic 
quantum dots.  In Sec.~\ref{applications} we test the derived functional 
for a large set of QDs and quantum rings. A summary is given in Sec.~\ref{summary}. 
%%%%%%%%%%%%%%%%%%%%%%%%%%%%%%%%%%%%%%%%%%%%%%%%%%%%%%%%%%%%%%%%%%%%%%%

%%%%%%%%%%%%%%%%%%%%%%%%%%%%%%%%%%%%%%%%%%%%%%%%%%%%%%%%%%%%%%%%%%%%%%%
%%%%%%%%%%%%%%%%%%%%%%%%%%%%%%%%%%%%%%%%%%%%%%%%%%%%%%%%%%%%%%%%%%%%%%%
\section{Construction of B88 in two dimensions}\label{theory}
%%%%%%%%%%%%%%%%%%%%%%%%%%%%%%%%%%%%%%%%%%%%%%%%%%%%%%%%%%%%%%%%%%%%%%%
%%%%%%%%%%%%%%%%%%%%%%%%%%%%%%%%%%%%%%%%%%%%%%%%%%%%%%%%%%%%%%%%%%%%%%%

%%%%%%%%%%%%%%%%%%%%%%%%%%%%%%%%%%%%%%%%%%%%%%%%%%%%%%%%%%%%%%%%%%%%%%%
\subsection{General considerations}
%%%%%%%%%%%%%%%%%%%%%%%%%%%%%%%%%%%%%%%%%%%%%%%%%%%%%%%%%%%%%%%%%%%%%%%
The functional known as B88 in 3D is -- as a fundamental ingredient of 
B3LYP~\cite{b3lyp} -- among the most popular density functionals. It 
defines the energy density per electron with appealing features for 
finite systems such as a proper tail,~\cite{b88} and it recovers 
an appropriate small-gradient limit.~\cite{b88,elliott}

Let us first briefly remind of the B88 expression in 3D. For (globally 
collinear) spin-polarized states, it is convenient to write the exchange 
energy in terms of the exchange-hole potential $U_{{\rm X}, \sigma}$ as
\be\label{B88}
 E^{\rm B88}_{\rm X} = \frac{1}{2} \sum_\sigma \int d^3 r~ n_\sigma({\bf r})  U^{\rm B88}_{{\rm X}, \sigma}({\bf r})
\ee
and split it into two contributions
\be\label{B88b}
U^{\rm B88}_{{\rm X}, \sigma}({\bf r})  = U^{\rm LDA}_{{\rm X}, \sigma}({\bf r})  + \Delta U^{\rm B88}_{{\rm X}, \sigma}({\bf r}).
\ee
Here the first term comes from the LDA,
\be\label{B88c}
U^{\rm LDA}_{{\rm X},\sigma}  =  C_{\rm X} n^{1/3}_\sigma,~~ C_{\rm X} =  - 3 \left[ \frac{3}{4 \pi} \right]^{1/3},
\ee
and the second term is introduced in order to account for the inhomogeneities of the system through an expression
\be\label{B88d}
\Delta U^{\rm B88}_{{\rm X}, \sigma} =  -\beta \frac{n^{1/3}_\sigma x^2_\sigma}{1 + 6 \beta x_\sigma \sinh^{-1} ( x_\sigma )},
\ee
that depends on the dimensionless gradient  
\be\label{x3d}
x_\sigma=\frac{|\nabla n_\sigma|}{n^{4/3}_\sigma}\;.
\ee

A straightforward dimensional analysis suggests the following 2D version 
\be\label{2D-B88a}
 E^{\rm 2D-B88}_{\rm X} = \frac{1}{2} \sum_\sigma \int d^2 r~ n_\sigma({\bf r})  U^{\rm 2D-B88}_{{\rm X}, \sigma}({\bf r}),
\ee
where
\be\label{2D-DB88b}
U^{\rm 2D-B88}_{{\rm X}, \sigma}({\bf r})  = U^{\rm LDA}_{{\rm X}, \sigma}({\bf r})  + \Delta U^{\rm 2D-B88}_{{\rm X}, \sigma}({\bf r}) \;,
\ee
with
\be
U^{\rm 2D-LDA}_{{\rm X},\sigma}  = C^{\rm 2D}_{\rm X}  n^{1/2}_\sigma\;,
\ee
\be
\Delta U^{\rm 2D-B88}_{{\rm X}, \sigma} =  -\beta_{\rm 2D} \frac{ n^{1/2}_\sigma \tilde{x}^2_\sigma}{1 + \gamma \beta_{\rm 2D} ~ \tilde{x}_\sigma \sinh^{-1} ( \tilde{x}_\sigma )}\;,
\ee
and the 2D dimensionless gradient
\be\label{x}
\tilde{x}_\sigma=\frac{|\nabla n_\sigma|}{n^{3/2}_\sigma}\;.
\ee

In this case, however, the dimensional analysis cannot determine the 
coefficients $C^{\rm 2D}_{\rm X}$, $\gamma$, and $\beta_{\rm 2D}$.For 
$C^{\rm 2D}_{\rm X}$ it is tempting to use the value provided by the 
2D-LDA~\cite{rajagopal}: $C^{\rm 2D}_{\rm X} = - 16/(3 \sqrt{\pi})$ 
(this choice will be further justified below). In order to determine
$\gamma$, we require that the  2D exchange-hole potential behaves as
$-1/r$ at large $r$ (Ref.~\cite{2d9}) for densities that behave as 
$e^{-a_\sigma r^2}$, which is the case in, e.g., parabolic QDs. In this
way we obtain $\gamma = 8$.

As the last step, we need to find $\beta_{\rm 2D}$, where we start by 
observing that $\beta_{\rm 2D}$ would provide  the coefficient of the 
quadratic term of the small-gradient limit, i.e.,
\be\label{SmallX}
\Delta  E^{\rm 2D-B88}_{\rm X}  \approx -\beta_{2D} \sum_\sigma \int d^2 r~  n^{3/2}_\sigma \tilde{x}^2_\sigma\;.
\ee
However, standard techniques applied to obtain a gradient expansion in
2D fail to yield finite coefficients.~\cite{kirzhnits} Therefore, it is
crucial to look for alternative ways to define the gradient expansion
in some proper sense.
%%%%%%%%%%%%%%%%%%%%%%%%%%%%%%%%%%%%%%%%%%%%%%%%%%%%%%%%%%%%%%%%%%%%%%%

%%%%%%%%%%%%%%%%%%%%%%%%%%%%%%%%%%%%%%%%%%%%%%%%%%%%%%%%%%%%%%%%%%%%%%%
\subsection{Semiclassical limit via scaling of the potential and particle number}
%%%%%%%%%%%%%%%%%%%%%%%%%%%%%%%%%%%%%%%%%%%%%%%%%%%%%%%%%%%%%%%%%%%%%%%
In 3D, $\beta$ was obtained by Becke through the fitting of 
(Hartree-Fock) exchange energies for ``high-Z'' noble-gas atoms.~\cite{b88}
Recently, Elliott and Burke proved  that this choice has a fully 
non-empirical character.~\cite{elliott} In particular, they elucidated
-- through a careful and accurate numerical analysis at the level of 
exact-exchange (EXX) calculations -- that, in the high-Z limit, the 
local exchange gives the leading contribution to the exchange 
energy,~\cite{S81,FS90} and the second-order gradient corrections yields
the leading contribution of local inhomogeneities with a coefficient 
very close to the one found by Becke.~\cite{b88} This coefficient is 
different from the one that may be deduced from standard gradient 
expansions, where a weakly inhomogeneous extended periodic system is 
used as the reference. {\em Finite} systems cannot be considered weakly
inhomogeneous,  but their high-Z limit corresponds to a favorable 
exception~\cite{PBEsol,elliott} emerging from the exact behavior of 
interacting quantum systems~\cite{LS73,LS77,S81,FS90}.

Next, we show that a similar idea and procedure applies to 2D. We 
restrict the analysis to parabolic quantum dots, often referred as 
artificial atoms of the 2D world. Lieb and co-workers~\cite{lieb2D}
have rigorously proven that if the constant $\omega$ of a parabolic 
confinement potential
\be
V_{\rm ext}(r) = \frac{1}{2} m \omega^2\,r^2
\label{conf}
\ee
is scaled with the particle number as
\be\label{2DLS}
N \rightarrow  N' = \lambda N,~~ \omega \rightarrow  \omega' = \sqrt{\lambda} \omega\;,
\ee
the 2D Thomas-Fermi (TF) theory provides the leading contribution to 
the {\em total} energy for large $N$ (Ref.~\cite{lieb2D}). 
Correspondingly, the TF density, $n_{\rm TF}$ will reproduce the exact
density in an averaged sense. In other words, the system becomes 
increasingly semiclassical as a function of $N$. In the following, we
explore the situation at the level of exchange. All the numerical 
results were obtained with the code {\sc octopus}~\cite{octopus}.

\begin{figure}
\includegraphics[width=\columnwidth]{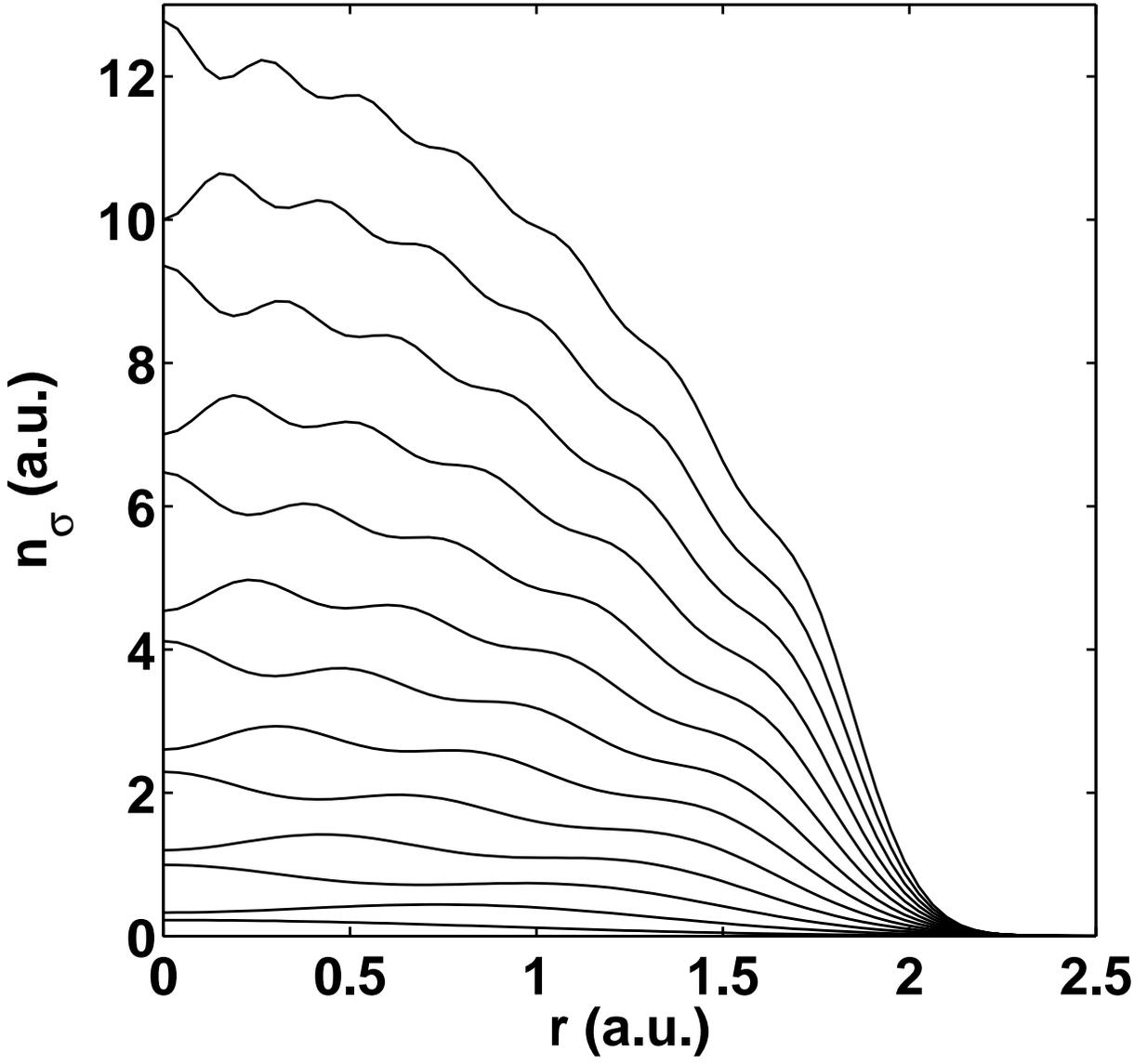}
\caption{Electron densities for parabolic quantum dots with $N=2,\ldots,182$ 
electrons scaled according to Eq.~(\ref{2DLS}). 
As a function of $N$ (from bottom to top), the spatial extent is preserved 
and the relative density oscillations become smaller.}
\label{fig1}
\end{figure}

\begin{figure}
\includegraphics[width=\columnwidth]{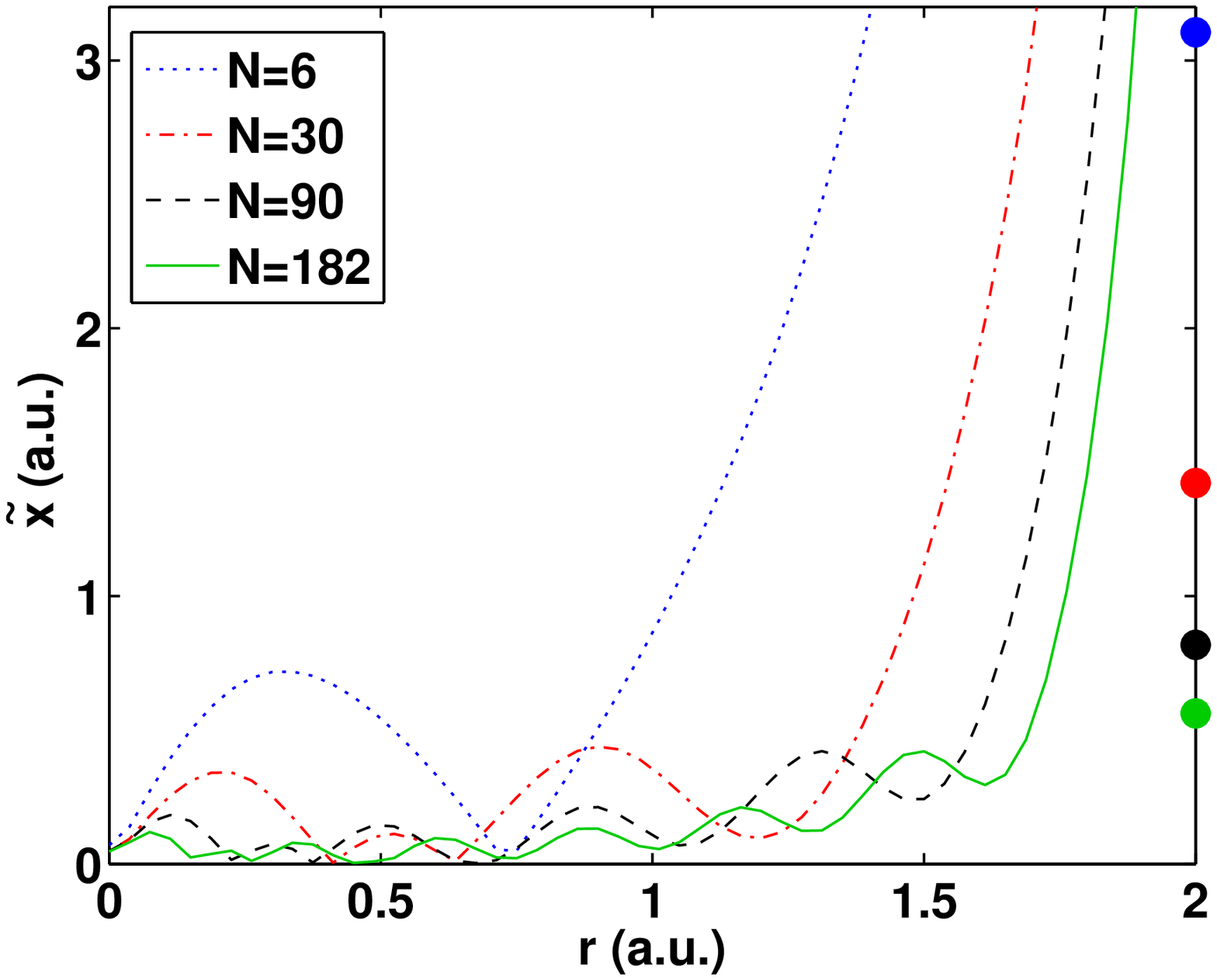}
\caption{Dimensionless gradient ${\tilde x}$ for (scaled) parabolic quantum
dots with $N=6,\,30,\,90,\,182$, respectively (cf. the densities
in \ref{fig1}). The circles on the right correspond to
the mean values of $x$ in the range $r=0 \ldots 2$ a.u.}
\label{fig2}
\end{figure}

In \ref{fig1} we show the electronic density of a series of 
{\em closed-shell} parabolic QDs ($N=2,6,12,\ldots,182$) obeying 
Eq.~(\ref{2DLS}) with initial $\omega_{N=2} = 1$ (a.u.). Clearly the 
density increases with $N$ while its radial extent remains approximately
constant. The picture suggests that the system is gradually approaching 
the high-density limit. Consequently, exchange effects will eventually 
dominate over correlation. Moreover, the {\em relative} amplitude of 
the density oscillations gradually become negligible. This is evident 
in \ref{fig2} that shows the corresponding dimensionless gradients 
$\tilde{x}$. It is appealing to conclude that, asymptotically, the LDA
provides the ``exact'' result for exchange -- as the region of the 
divergence of $\tilde{x}$ becomes energetically irrelevant. This is 
clarified further below.

\begin{figure}
\includegraphics[width=\columnwidth]{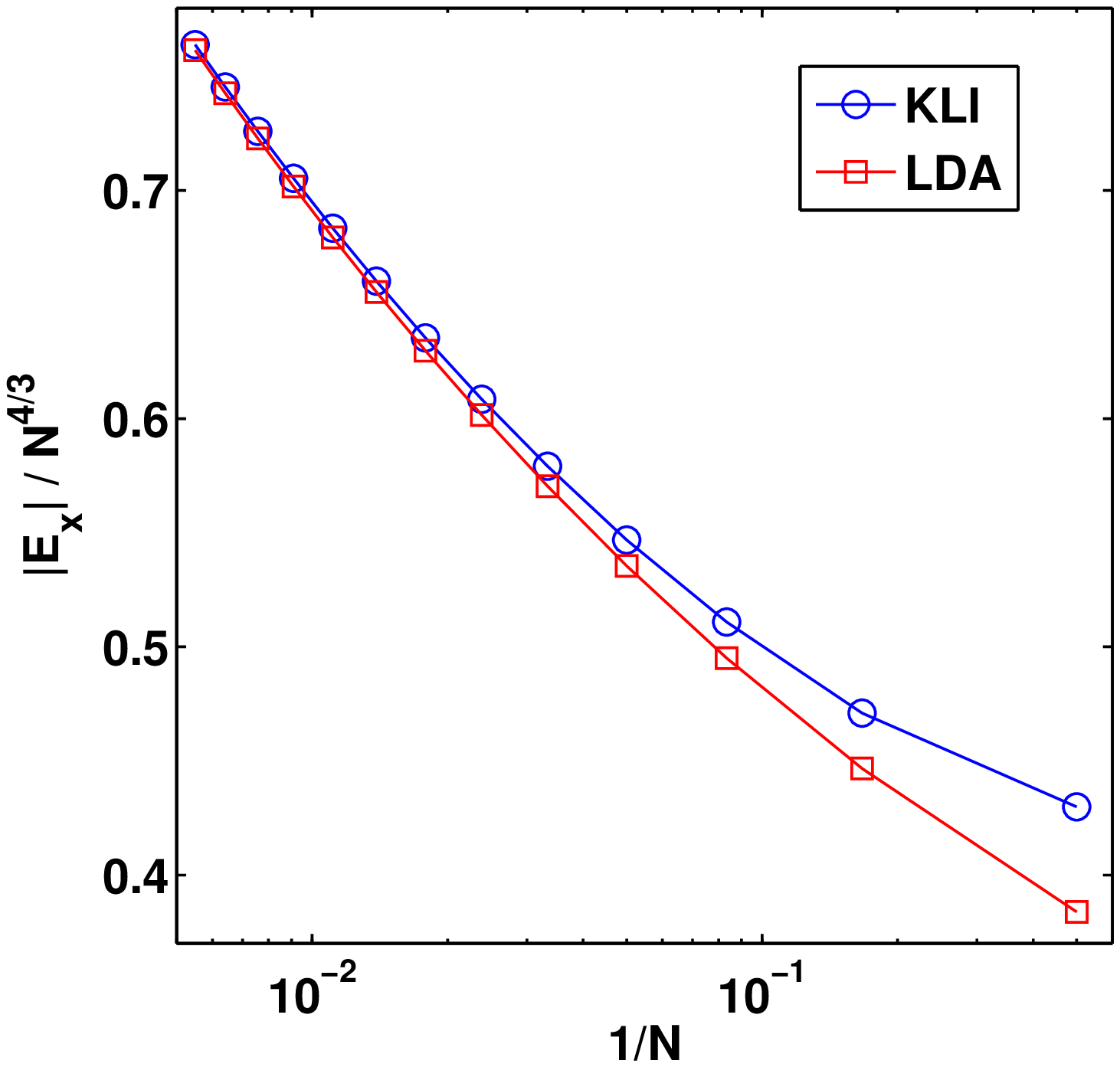}
\caption{Comparison of exchange energies of the exact-exchange 
scheme in the KLI approximation and the two-dimensional local-density 
approximation (LDA) for a set of parabolic quantum dots at the order of $N^{4/3}$
(see text). The LDA results approach the KLI values as a function of $N$. }
\label{fig3}
\end{figure}

The density satisfies asymptotically the scaling relation 
\be\label{ScalingD} n_{{\rm TF}, N}({\bf r}) = N n_{{\rm TF}, 1}({\bf r})\;.
\ee Using Eq.~(\ref{ScalingD}) in Eqs.~(\ref{B88c}) and (\ref{SmallX})
we find that the LDA exchange energies are of the order $N^{4/3}$, and
the second-order gradient corrections are of the order $N^{1/2}$,
respectively. \ref{fig3} shows that the LDA and exact exchange
(EXX) energies, the latter evaluated within the
Krieger-Lee-Iafrate~\cite{kli} (KLI) approximation, converge to the
same value at the order $N^{4/3}$.  This analysis justifies using the
value $C^{\rm 2D}_{\rm X} = - 16/(3 \sqrt{\pi})$ that stems from the
2D-LDA.

\begin{figure}
\includegraphics[width=\columnwidth]{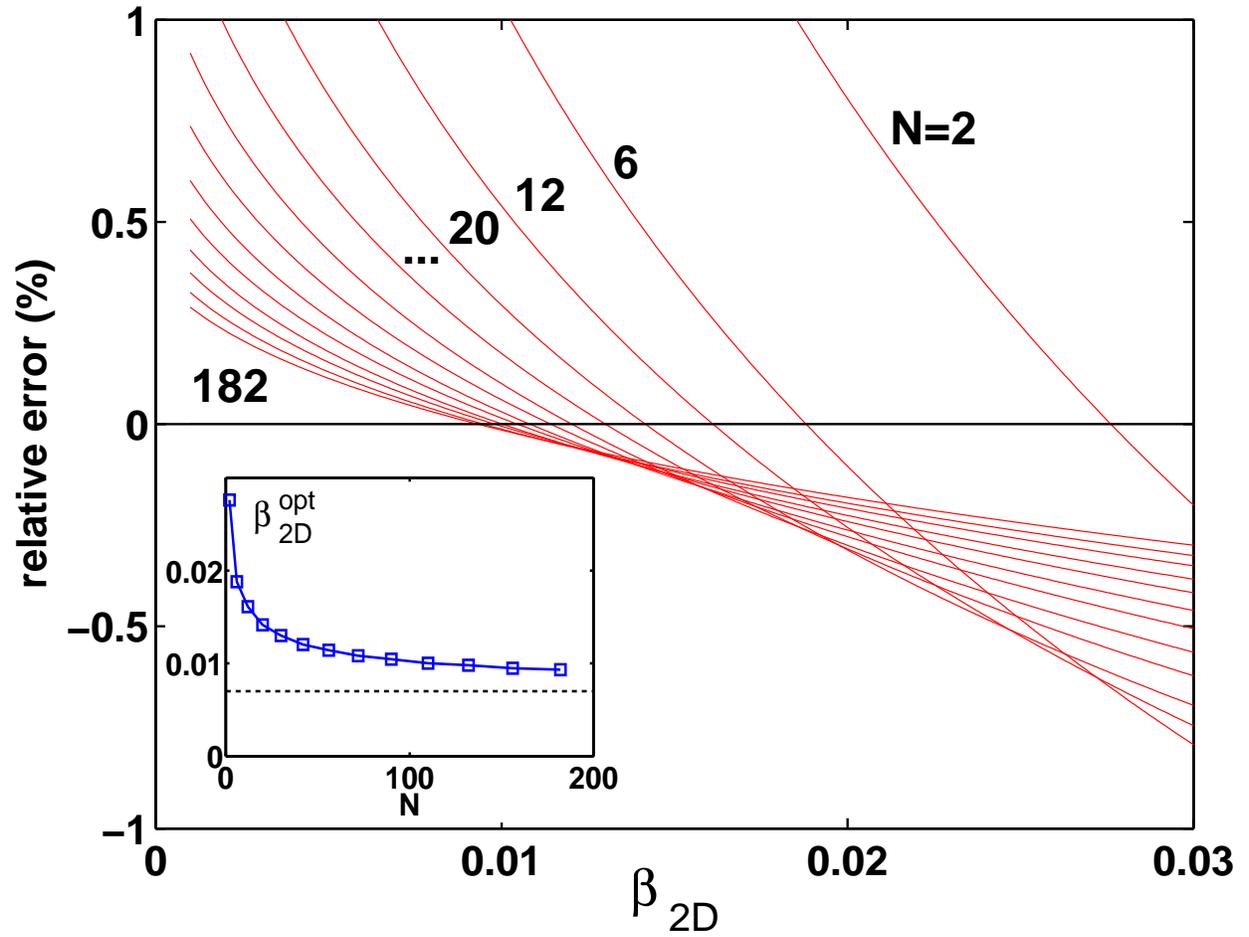}
\caption{Relative error of the 2D-B88 functional
with respect to the EXX-KLI results (see text) 
as a function of $\beta_{\rm 2D}$ for parabolic quantum dots with 
$N=2,\ldots,182$.  The 
optimal $\beta_{\rm 2D}$ (zero error) as a function of $N$
are seen to converge in the inset. The horizontal
line in the inset shows our choice $\beta_{\rm 2D} \rightarrow 0.007$.}
\label{fig4}
\end{figure}

Now we proceed to the next order, $N^{1/2}$, and try to determine 
$\beta_{\rm 2D}$ numerically. \ref{fig4} shows the relative 
error of the 2D-B88 functional as a function of $\beta_{\rm 2D}$ 
for the same set of parabolic quantum dots with $N=2,6,12,\ldots,182$. 
For each $N$, we determine the optimal $\beta_{\rm 2D}$ that gives 
zero error. The inset of \ref{fig4} shows the behavior of this 
sequence. A simple polynomial fit leads to $\beta_{\rm 2D}=0.007$ in 
the $N\rightarrow\infty$ limit. We point out, however, that there is 
uncertainty in this value beacause of the following reasons. First, our
analysis is limited by $N_{\rm max}=182$ due to the demanding 
convergence of the EXX-KLI reference results on a cartesian grid.
This prevents us to fully explore the asymptotic region, and we cannot 
exclude the possibility of small numerical errors at the order of 
$N^{4/3}$ affecting the estimation made at next (lower) order in $N$. 
Secondly, the full optimized-effective-potential (OEP) scheme might 
lead to a different optimal value of $\beta_{\rm 2D}$, although it is 
known (in the 3D case) that the KLI approximation is typically very 
close to the OEP result. Thirdly, different confinement/boundary 
conditions (i.e., different types of turning points) can, in principle,
lead to different (yet fully non-empirical) optimal values for 
$\beta_{\rm 2D}$. Therefore, having a different reference system 
instead of a parabolic quantum dot could affect the result as well; 
this aspect is touched in the next section, where we explore the
performance of our functional on quantum rings.

Despite the uncertainties listed above, the general principles in the
determination of $\beta_{\rm 2D}$ are clear. Therefore we proceed by 
choosing $\beta_{\rm 2D}=0.007$, and assess the performance of the 
functional in detail in the following section.

%%%%%%%%%%%%%%%%%%%%%%%%%%%%%%%%%%%%%%%%%%%%%%%%%%%%%%%%%%%%%%%%%%%%%%%

%%%%%%%%%%%%%%%%%%%%%%%%%%%%%%%%%%%%%%%%%%%%%%%%%%%%%%%%%%%%%%%%%%%%%%%
\section{Performance in applications}\label{applications}
%%%%%%%%%%%%%%%%%%%%%%%%%%%%%%%%%%%%%%%%%%%%%%%%%%%%%%%%%%%%%%%%%%%%%%%
Next we test our 2D-B88 functional {\em self-consistently} for realistic
2D systems in comparison with exchange-only energies obtained with the 
KLI and local-density approximations. We shall take a look at systems not 
included in the estimation of $\beta_{\rm 2D}$. \ref{table1} shows 
the exchange energies of parabolic QDs with various $N$ and confinement
strengths $\omega$ [see Eq.~(\ref{conf})]. The relative errors of the 
approximations are given in the last two columns. Overall, we find 
excellent agreement between 2D-B88 and EXX-KLI with a mean relative 
error of $1.7\%$ for the whole set. In comparison, the 2D-LDA yields a 
mean error of $5.2\%$. We note that, as expected, both approximations 
improve their accuracy as a function of $N$.

% =====================   BEGIN TABLE  ============================= %
\begin{table}
  \caption{\label{table1}
Exchange energies (in atomic units) for parabolic quantum dots with
varying $N$ and confinement strength $\omega$.
The columns correspond to the exact exchange in the KLI approximation, 
the local density-approximation (LDA), and the
generalized-gradient approximation presented in this work (2D-B88).
}
  \begin{tabular}{c c c c c c c}
  \hline
  \hline 
  $N$ & $\omega$ & $-E^{\rm KLI}_{\rm x}$ &  $-E_{\rm x}^{\rm LDA}$ & $-E_{\rm x}^{\rm 2D-B88}$ & $|\Delta_{\rm rel}^{\rm LDA}|$  & $|\Delta_{\rm rel}^{\rm 2D-B88}|$ \\
  \hline
    2 & 0.5 &      0.7291 &      0.6495  &      0.6992  &   10.9$\%$ &  4.10$\%$\\
    2 & 1.5 &      1.3583 &      1.2147  &      1.3048  &   10.6$\%$  & 3.94$\%$\\
    2 & 2.5 &      1.7979 &      1.6106  &      1.7284  &   10.4$\%$  & 3.87$\%$\\
    2 & 3.5 &      2.1571 &      1.9343  &      2.0745  &   10.3$\%$  & 3.83$\%$\\
    6 & 0.5 &      2.4707 &      2.3392  &      2.4311   &  5.32$\%$  & 1.60$\%$\\
    6 & 1.5 &      4.7267 &      4.4823  &      4.6486  &   5.17$\%$  & 1.65$\%$\\
    6 & 2.5 &      6.3311 &      6.0081  &      6.2266  &   5.10$\%$  & 1.65$\%$\\
    6 & 3.5 &      7.6509 &      7.2638  &      7.5252 &    5.06$\%$  & 1.64$\%$\\
   12 & 0.5 &      5.4316 &      5.2571  &      5.3875  &   3.21$\%$  & 0.81$\%$\\
   12 & 1.5 &     10.535  &     10.206  &     10.444  &      3.13$\%$  & 0.87$\%$\\
   12 & 2.5 &     14.204  &     13.765  &     14.080  &      3.09$\%$  & 0.88$\%$\\
   12 & 3.5 &     17.237  &     16.709  &     17.086  &      3.06$\%$  & 0.88$\%$\\
   20 & 0.5 &      9.7651 &      9.5537  &      9.7229  &      2.16$\%$  & 0.43$\%$\\
   20 & 1.5 &     19.107  &     18.704  &     19.013  &      2.11$\%$  & 0.49$\%$\\
   20 & 2.5 &     25.874  &     25.334  &     25.744  &      2.09$\%$  & 0.50$\%$\\
   20 & 3.5 &     31.490  &     30.837  &     31.330  &      2.07$\%$ &  0.51$\%$\\
  \hline
  \multicolumn{3}{l}{mean error} &   &  & 5.2$\%$ & 1.7$\%$  \\
  \hline
  \hline
  \end{tabular}
\end{table}
% ========================   END TABLE  ============================= %

In \ref{table2} we examine the performance of the 2D-B88 
in QDs at low electron densities (small confinement strengths for
only a few electrons).
This regime is important in view of QD applications exploiting
strongly correlated electrons. Again, we find that 2D-B88 clearly
overperforms the LDA and yields very accurate exchange energies
in comparison with the EXX-KLI. However, it remains to be tested
how the 2D-B88 works in combination with a carefully chosen functional
for the correlation. Only such a combined functional would be truly 
useful for applications in the low-density regime. 

\ref{table3} shows exchange energies for fully spin-polarized
($S=N/2$) parabolic quantum dots with a relatively low confinement
strength. As in the previous examples, 2D-B88 is very accurate. This 
test validates the usability of the functional in a fully spin-dependent 
fashion according to the above formulation.

% =====================   BEGIN TABLE  ============================= %
\begin{table}
  \caption{\label{table2} Exchange energies (in atomic units) 
  for low-density parabolic quantum dots. 
The columns correspond to the exact exchange in the KLI approximation, 
the local density-approximation (LDA), and the
generalized-gradient approximation presented in this work (2D-B88).
}
  \begin{tabular}{c c c c c c c}
  \hline
  \hline
  $N$ & $\omega$ & $-E^{\rm KLI}_{\rm x}$ &  $-E_{\rm x}^{\rm LDA}$ & $-E_{\rm x}^{\rm 2D-B88}$ & $|\Delta_{\rm rel}^{\rm LDA}|$  & $|\Delta_{\rm rel}^{\rm 2D-B88}|$ \\
  \hline
    2 & 1          & 1.0831 &      0.9673  &      1.0398 & 10.7$\%$ & 4.00$\%$\\
    2 & 1/4        & 0.4851 &      0.4312  &      0.4647 & 11.1$\%$ & 4.21$\%$ \\
    2 & 1/6        & 0.3801 &      0.3376  &      0.3640 & 11.2$\%$ & 4.24$\%$ \\
    2 & 1/16       & 0.2075 &      0.1844  &      0.1993 & 11.1$\%$ & 3.95$\%$ \\
    2 & 1/36       & 0.1275 &      0.1141  &      0.1268 & 10.5$\%$ & 0.55$\%$ \\
    6 & 1/4        & 1.6185 &      1.5312  &      1.5943 & 5.39$\%$ & 1.50$\%$ \\ 
    6 & 1/16       & 0.6766 &      0.6403  &      0.6697 & 5.37$\%$ & 1.02$\%$ \\
  \hline
  \multicolumn{3}{l}{mean error} & & & 9.3$\%$ & 2.8$\%$ \\
  \hline
  \hline
  \end{tabular}
\end{table}
% ========================   END TABLE  ============================= %

% =====================   BEGIN TABLE  ============================= %
\begin{table}
  \caption{\label{table3} Exchange energies (in
  atomic units) for spin-polarized ($S=N/2$) parabolic quantum dots. 
The columns correspond to the exact exchange in the KLI approximation, 
the local density-approximation (LDA), and the
generalized-gradient approximation presented in this work (2D-B88).
}
  \begin{tabular}{c c c c c c c c}
  \hline
  \hline
  $N$ & $\omega$ & $-E^{\rm KLI}_{\rm x}$ &  $-E_{\rm x}^{\rm LDA}$ & $-E_{\rm x}^{\rm 2D-B88}$ & $|\Delta_{\rm rel}^{\rm LDA}|$  & $|\Delta_{\rm rel}^{\rm 2D-B88}|$ \\
  \hline
    2 & 1/4  &    0.6645 &      0.6018  &      0.6421 & 9.43$\%$ & 3.37$\%$ \\ 
    3 & 1/4  &    1.0146 &      0.9533  &      0.9987 & 6.04$\%$ & 1.57$\%$ \\
    4 & 1/4  &    1.4303 &      1.3363  &      1.4019 & 6.57$\%$ & 1.99$\%$ \\
    5 & 1/4  &   1.8091 &      1.7228  &      1.7876 & 4.77$\%$ & 1.19$\%$ \\
    6 & 1/4  &    2.1973 &      2.1177  &      2.1813 & 3.62$\%$ & 0.73$\%$ \\
    2 & 1/16 &    0.3182 &      0.2765  &      0.3035 & 13.1$\%$ & 4.62$\%$ \\
    3 & 1/16 &   0.4607 &      0.4296  &      0.4631 & 6.75$\%$ & 0.52$\%$ \\
    4 & 1/16 &    0.6697 &      0.5979  &      0.6487 & 10.7$\%$ & 3.14$\%$ \\
    5 & 1/16 &    0.8165 &      0.7607  &      0.8064 & 6.83$\%$ & 1.24$\%$ \\
    6 & 1/16 &    0.9709 &      0.9265  &      0.9853 & 4.57$\%$ & 1.48$\%$ \\
  \hline
  \multicolumn{4}{l}{mean error} &  & 7.2\% & 2.0\%     \\
  \hline
  \hline
  \end{tabular}
\end{table}
% ========================   END TABLE  ============================= %

\begin{figure}
\includegraphics[width=0.68\columnwidth]{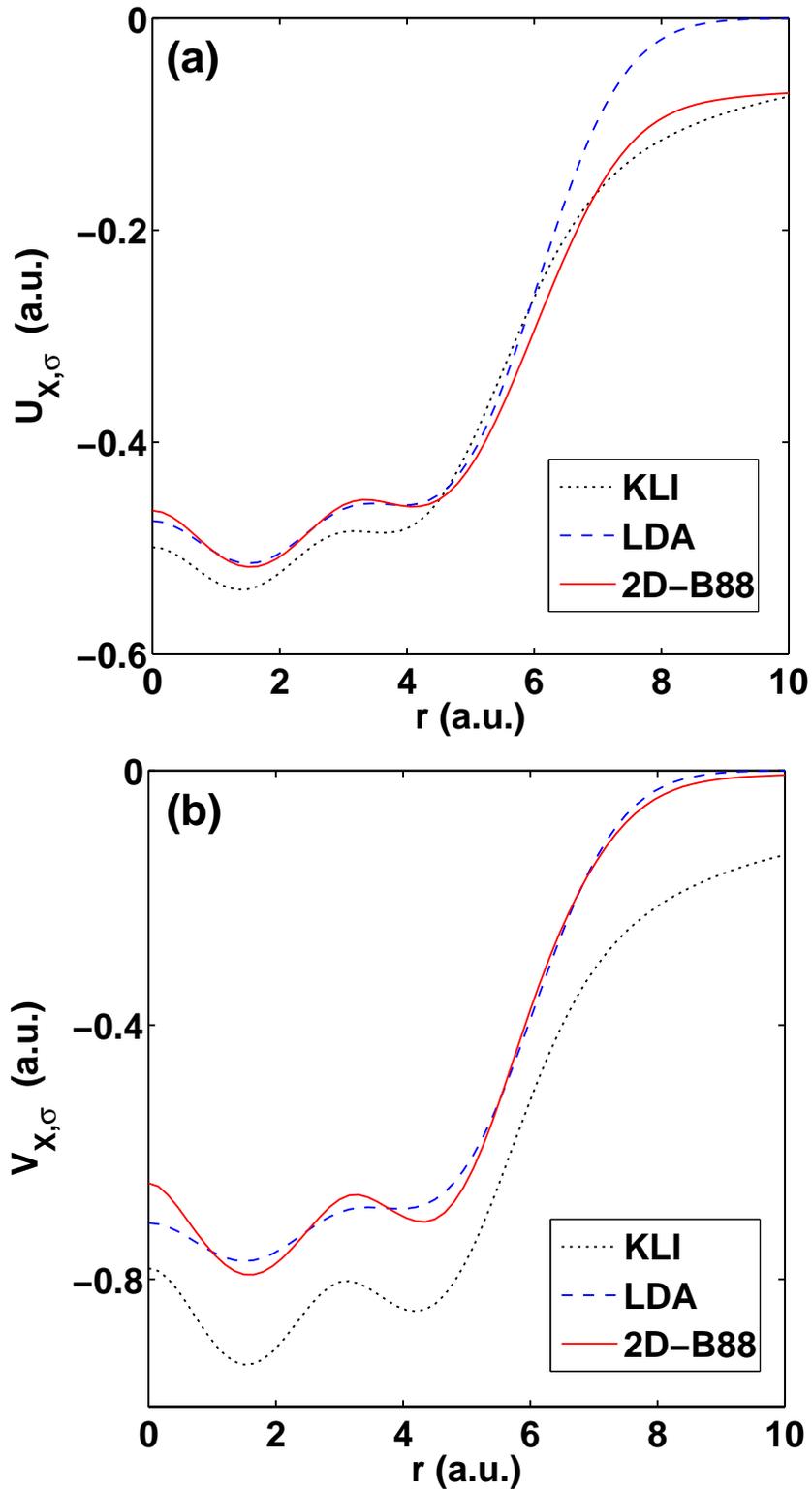}
\caption{Self-consistent exchange-hole potentials (a) and Kohn-Sham exchange
potentials (b) calculated with EXX-KLI, 2D-LDA, and 2D-B88, 
respectively (exchange only) for a 20-electron parabolic quantum
dot with $\omega=0.4217$ a.u.}
\label{fig5}
\end{figure}

Besides the exchange energies, it is informative to compare the 
exchange-hole potentials as well as the Kohn-Sham exchange potentials.
\ref{fig5}(a) shows $U_{{\rm X},\sigma}$ for an $N=20$ 
parabolic QD with $\omega=0.4217$ a.u. The structure of the potential
in the central part (within the shells) is very similar in the LDA 
and 2D-B88. However, the latter functional is able to describe the 
asymptotic behavior very accurately in comparison with the EXX-KLI 
result. In turn, this  leads to accurate exchange energies given by
2D-B88. The behavior at large $r$ is expected due to the form in 
Eq.~(\ref{B88d}) that  resembles the asymptotically corrected functional
for the exchange-correlation potential in Ref.~\cite{2d13}.

The Kohn-Sham exchange potentials $V_{\rm X}$ are shown in 
\ref{fig5}(b). In this case, neither LDA nor 2D-B88 are able to 
give the correct asymptotic behavior. However, it is interesting to see
that the 2D-B88 potential produces the shell structure more accurately 
than the LDA at $0 < r \lesssim 5$ a.u. We also find close similarity in
the shell region between the 2D-B88 and the meta-GGA result suggested 
in Ref.~\cite{2d9} for $V_{\rm X}$. We note, however, that the values 
in the lower panel of Fig. 4 in that reference miss a factor of two.

% =====================   BEGIN TABLE  ============================= %
\begin{table}
  \caption{\label{table4} Exchange energies (in
  atomic units) for a quantum ring (see text). 
The columns correspond to the exact exchange in the KLI approximation, 
the local density-approximation (LDA), and the
generalized-gradient approximation presented in this work (2D-B88).
}
  \begin{tabular}{c c c c c c}
  \hline
  \hline  
  $N$ &  $-E^{\rm KLI}_{\rm x}$ &  $-E_{\rm x}^{\rm LDA}$ &  $-E_{\rm x}^{\rm 2D-B88}$ & $|\Delta_{\rm rel}^{\rm LDA}|$  & $|\Delta_{\rm rel}^{\rm 2D-B88}|$ \\
  \hline
6  &  2.1590 &  2.1095 &  2.2668 & 2.29$\%$ &  4.99$\%$\\
10 &  4.5192 &  4.3106 &  4.5458 & 4.62$\%$ &  0.59$\%$\\
14 &  7.1495 &  6.7915 &  7.0867 & 5.01$\%$ &  0.88$\%$\\
20 & 10.820 & 10.568 & 10.883 & 2.33$\%$ &  0.58$\%$\\
24 & 13.356 & 13.126 & 13.437 & 1.72$\%$ &  0.61$\%$\\
  \hline
  \multicolumn{2}{l}{mean error} & & & 3.2\% & 1.5\%   \\
  \hline
  \hline
  \end{tabular}
\end{table}
% ========================   END TABLE  ============================= %

For the usefulness of the 2D-B88 functional it is important to test its
validity for different physical systems. 
In the following we open the discussion to this direction by examining 
a quantum ring. As the system has a different topology from a QD it 
gives useful insights into the general applicability of the functional.
The external potential is now defined as $V_{\rm ext}(r)=\omega^2 (r-r_0)^2/2$, 
where we set $\omega=1$ a.u. and $r_0=3$ a.u. The results for exchange
energies are given in \ref{table4}. Overall, 2D-B88 yields 
significantly more accurate results than the LDA except for the 
$N=6$ ring. In that particular case the width of the electron density
along the perimeter is relatively small -- approaching the 
quasi-one-dimensional system, which calls for more elaborate ways
to deal with the electronic exchange.~\cite{2d6}
Nevertheless, at $N>6$ the accuracy of the 2D-B88 is excellent: the
deviation from the KLI exchange energy remains below $1\%$. 
%%%%%%%%%%%%%%%%%%%%%%%%%%%%%%%%%%%%%%%%%%%%%%%%%%%%%%%%%%%%%%%%%%%%%%%
%%%%%%%%%%%%%%%%%%%%%%%%%%%%%%%%%%%%%%%%%%%%%%%%%%%%%%%%%%%%%%%%%%%%%%%
%%%%%%%%%%%%%%%%%%%%%%%%%%%%%%%%%%%%%%%%%%%%%%%%%%%%%%%%%%%%%%%%%%%%%%%

%%%%%%%%%%%%%%%%%%%%%%%%%%%%%%%%%%%%%%%%%%%%%%%%%%%%%%%%%%%%%%%%%%%%%%%
%%%%%%%%%%%%%%%%%%%%%%%%%%%%%%%%%%%%%%%%%%%%%%%%%%%%%%%%%%%%%%%%%%%%%%%
\section{Summary}\label{summary}
%%%%%%%%%%%%%%%%%%%%%%%%%%%%%%%%%%%%%%%%%%%%%%%%%%%%%%%%%%%%%%%%%%%%%%%
%%%%%%%%%%%%%%%%%%%%%%%%%%%%%%%%%%%%%%%%%%%%%%%%%%%%%%%%%%%%%%%%%%%%%%%
In this work we constructed a generalized-gradient approximation for
the exchange energy in two dimensions (2D). With the construction we
overcame the known problems in finding finite coefficients for the 2D
gradient expansion through, e.g., the Kirzhnits expansion. Our
formulation follows the B88 exchange functional. The final coefficient
was then found through a fitting to properly scaled 2D harmonic
oscillators in the large-$N$ limit, corresponding to the high-Z limit
in three-dimensional atomic systems.  We tested the obtained
exchange-energy functional for various quantum dots and found
excellent agreement with exact-exchange results and a significant
improvement over the standard local-density approximation.  The
functional also leads to a proper asymptotic tail of the exchange-hole
potential and a more accurate exchange potential than that of the
local-density approximation. The generality of the functional was
confirmed in tests for low-density quantum dots, spin-polarized
systems, as well as 2D quantum rings. Possible further extensions 
of the present construction could include adaptation to the recently 
developed density-functional formalism for strongly interacting 
electrons.~\cite{paola1,paola2}
%%%%%%%%%%%%%%%%%%%%%%%%%%%%%%%%%%%%%%%%%%%%%%%%%%%%%%%%%%%%%%%%%%%%%%%
%%%%%%%%%%%%%%%%%%%%%%%%%%%%%%%%%%%%%%%%%%%%%%%%%%%%%%%%%%%%%%%%%%%%%%%
%%%%%%%%%%%%%%%%%%%%%%%%%%%%%%%%%%%%%%%%%%%%%%%%%%%%%%%%%%%%%%%%%%%%%%%

%%%%%%%%%%%%%%%%%%%%%%%%%%%%%%%%%%%%%%%%%%%%%%%%%%%%%%%%%%%%%%%%%%%%%%%
%%%%%%%%%%%%%%%%%%%%%%%%%%%%%%%%%%%%%%%%%%%%%%%%%%%%%%%%%%%%%%%%%%%%%%%
% The "Acknowledgement" section 
%%%%%%%%%%%%%%%%%%%%%%%%%%%%%%%%%%%%%%%%%%%%%%%%%%%%%%%%%%%%%%%%%%%%%%%
%%%%%%%%%%%%%%%%%%%%%%%%%%%%%%%%%%%%%%%%%%%%%%%%%%%%%%%%%%%%%%%%%%%%%%%
\acknowledgement
We thank Alexander Odriazola for numerical support and useful
discussions. The work was supported by the 
European Community s FP7 through the CRONOS project, grant 
agreement no. 280879, the Academy of Finland through project 
no. 126205, and COST Action CM1204 XLIC.
S.P. also acknowledges support from NSF Grant CHE-1112442 and 
discussions with Kieron Burke. J.G.V. acknowledges support from the
FCT Grant No.SFRH/BD/38340/2007.
%%%%%%%%%%%%%%%%%%%%%%%%%%%%%%%%%%%%%%%%%%%%%%%%%%%%%%%%%%%%%%%%%%%%%%%
%%%%%%%%%%%%%%%%%%%%%%%%%%%%%%%%%%%%%%%%%%%%%%%%%%%%%%%%%%%%%%%%%%%%%%%
%%%%%%%%%%%%%%%%%%%%%%%%%%%%%%%%%%%%%%%%%%%%%%%%%%%%%%%%%%%%%%%%%%%%%%%

%%%%%%%%%%%%%%%%%%%%%%%%%%%%%%%%%%%%%%%%%%%%%%%%%%%%%%%%%%%%%%%%%%%%%%%
%%%%%%%%%%%%%%%%%%%%%%%%%%%%%%%%%%%%%%%%%%%%%%%%%%%%%%%%%%%%%%%%%%%%%%%

%%%%%%%%%%%%%%%%%%%%%%%%%%%%%%%%%%%%%%%%%%%%%%%%%%%%%%%%%%%%%%%%%%%%%%%
%%%%%%%%%%%%%%%%%%%%%%%%%%%%%%%%%%%%%%%%%%%%%%%%%%%%%%%%%%%%%%%%%%%%%%%
%%%%%%%%%%%%%%%%%%%%%%%%%%%%%%%%%%%%%%%%%%%%%%%%%%%%%%%%%%%%%%%%%%%%%%%


\begin{thebibliography}{ll}
%%%%%%%%%%%%%%%%%%%%%%%%%%%%%%%%%%%%%%%%%%%%%%%%%%%%%%%%%%%%%%%%%%%%%%%
%%%%%%%%%%%%%%%%%%%%%%%%%%%%%%%%%%%%%%%%%%%%%%%%%%%%%%%%%%%%%%%%%%%%%%%
\bibitem{qd} 
S. M. Reimann and M. Manninen.
{ \it Rev. Mod. Phys. }
{\bf 2002}, {\it 74}, 1283 ;
%
L. P. Kouwenhoven, D. G. Austing and S. Tarucha.
{ \it Rep. Prog. Phys. }
{\bf 2001}, {\it 64}, 701.

\bibitem{agreview} 
For review, see M. Polini, F. Guinea, M. Lewenstein, H.C. Manoharan, and V. Pellegrini.
{ \it Nat. Nanotechnol.}
{\bf 2013 }, {\it 8}, 625.

\bibitem{dft3}
U. von Barth. 
{ \it Phys. Scripta}
{\bf 2004 }, {\it T109},9 .

\bibitem{dft1} 
{\it Density Functional Theory} ;
R. M. Dreizler and E. K. U. Gross;
Springer: Berlin, 1990.

\bibitem{dft2} 
{\it A Primer in Density Functional Theory};
C. Fiolhais, F. Nogueira, and M.\,A.\,L. Marques, Eds.;
Springer-Verlag: Berlin, 2003.

\bibitem{functionals} 
J. P. Perdew and S. Kurth;
Density Functionals for Non-relativistic Coulomb Systems in the New Century.
In {\it A Primer in Density Functional Theory}, first edition;
C. Fiolhais, F. Nogueira, and M.\,A.\,L. Marques, Eds.;
Springer-Verlag: Berlin, 2003; pp 1.
%
G. E. Scuseria and V. N. Staroverov;
Progress in the development of exchange-correlation functionals.
In {\it Theory and Applications of Computational Chemistry: The First Forty Years}, first edition;
C. E. Dykstra, G. Frenking, K. S. Kim, and G. E. Scuseria, Eds.;
Elsevier: Amsterdam, 2005; pp. 669.

\bibitem{kim}
Y.-H. Kim, I.-H. Lee, S. Nagaraja, J.-P. Leburton, R. Q. Hood, and R. M. Martin.
{ \it Phys. Rev. B }
{\bf 2000 }, {\it 61}, 5202.

\bibitem{pollack}
L. Pollack and J. P. Perdew.
{ \it J. Phys.: Condens. Matt.}
{\bf 2000 }, {\it 12}, 1239.

\bibitem{chiodo}
L. Chiodo, L. A. Constantin, E. Fabiano, and F. Della Sala.
{ \it Phys. Rev. Lett. }
{\bf 2012 }, {\it 108}, 126402.

\bibitem{rajagopal}
A. K. Rajagopal and J. C. Kimball.
{ \it Phys. Rev. B }
{\bf 1977 }, {\it 15}, 2819.

\bibitem{tanatar}
B. Tanatar and D. M. Ceperley.
{ \it Phys. Rev. B}
{\bf 1989 }, {\it 39}, 5005.

\bibitem{attaccalite}
C. Attaccalite, S. Moroni, P. Gori-Giorgi, and G. B. Bachelet.
{ \it Phys. Rev. Lett. }
{\bf 2002 }, {\it 88}, 256601.

\bibitem{2d1}
S. Pittalis, E. R\"as\"anen, N. Helbig, and E. K. U. Gross.
{ \it Phys. Rev. B }
{\bf 2007 }, {\it 76}, 235314.

\bibitem{2d4}
S. Pittalis, E. R\"as\"anen, J. G. Vilhena, and M. A. L. Marques.
{ \it Phys. Rev. A }
{\bf 2009 }, {\it 79}, 012503.

\bibitem{2d5}
S. Pittalis, E. R\"as\"anen, C. Proetto, and E. K. U. Gross.
{ \it Phys. Rev. B }
{\bf 2009 }, {\it 79}, 085316.

\bibitem{2d6}
E. R\"as\"anen, S. Pittalis, C. R. Proetto, and E. K. U. Gross.
{ \it Phys. Rev. B }
{\bf 2009 }, {\it 79}, 121305(R).

\bibitem{2d7}
S. Pittalis, E. R\"as\"anen, and E. K. U. Gross.
{ \it Phys. Rev. A }
{\bf 2009 }, {\it 80}, 032515.

\bibitem{2d12}
S. Pittalis and E. R\"as\"anen.
{ \it Phys. Rev. B }
{\bf 2010 }, {\it 82}, 165123.

\bibitem{kirzhnits}  
A. Putaja, E. R\"as\"anen, R. van Leeuwen, J. G. Vilhena, and M. A. L. Marques.
{ \it Phys. Rev. B }
{\bf 2012 }, {\it 85}, 165101.

\bibitem{elliott}
P. Elliot and K. Burke.
{ \it Can. J. Chemistry}
{\bf 2009 }, {\it 87}, 1485.

\bibitem{b88}
A.D. Becke.
{ \it Phys. Rev. A}
{\bf 1988 }, {\it 38}, 3098.

\bibitem{S81}
J. Schwinger.
{ \it Phys. Rev. A }
{\bf 1981 }, {\it 24}, 2353.

\bibitem{FS90}
C.L. Fefferman and C.L. Seco.
{ \it B. Am. Math. Soc. }
{\bf 1990 }, {\it 23}, 525.

\bibitem{2d9}
S. Pittalis, E. R\"as\"anen, and C. R. Proetto.
{ \it Phys. Rev. B }
{\bf 2010 }, {\it 81}, 115108.

\bibitem{b3lyp} 
A. D. Becke.
{ \it J. Chem. Phys.}
{\bf 1993 }, {\it 98}, 5648; 
%
C. Lee, W. Yang, and R. G. Parr.
{ \it Phys. Rev. B }
{\bf 1988 }, {\it 37}, 785;
%
P. Stephens, F. Devlin, C. Chabalowski, and M. Frisch.
{ \it J. Phys. Chem.}
{\bf 1994 }, {\it 98}, 11623.

\bibitem{PBEsol} 
J. P. Perdew, A. Ruzsinszky, G. I. Csonka, O. A. Vydrov, G. E. Scuseria, L. A. Constantin, X. Zhou, and K. Burke.
{ \it Phys. Rev. Lett.}
{\bf 2008 }, {\it 100}, 136406; 
{\bf 2009 }, {\it 102}, 039902(E).

\bibitem{LS73}
E. Lieb and B. Simon.
{ \it Phys. Rev. Lett.}
{\bf 1973 }, {\it 31}, 681.

\bibitem{LS77}
E. Lieb and B. Simon. 
{ \it Adv. Math. }
{\bf 1977 }, {\it 23}, 22.

\bibitem{lieb2D} 
E. H. Lieb, J. P. Solovej, and J. Yngvason.
{ \it Phys. Rev. B }
{\bf 1995 }, {\it 51}, 10646.

\bibitem{octopus} 
M.\,A.\,L. Marques, A. Castro, G.\,F. Bertsch, and A. Rubio.
{ \it Comput. Phys. Commun. }
{\bf 2003 }, {\it 151}, 60;
%
A. Castro, H. Appel, M. Oliveira, C. A. Rozzi, X. Andrade, F. Lorenzen, M. A. L. Marques, E. K. U. Gross, and A. Rubio.
{ \it Phys. Status Solidi B}
{\bf 2006 }, {\it 243}, 2465.

\bibitem{kli} 
J. B. Krieger, Y. Li, and G. J. Iafrate.
{ \it Phys. Rev. A }
{\bf 1992 }, {\it 45},  101.

\bibitem{2d13} 
S. Pittalis and E. R\"as\"anen.
{ \it Phys. Rev. B }
{\bf 2010 }, {\it 82}, 195124.

\bibitem{paola1}
P. Gori-Giorgi, M. Seidl, and G. Vignale.
{ \it Phys. Rev. Lett. }
{\bf 2009 }, {\it 103}, 166402.

\bibitem{paola2}
C. B. Mendl, F. Malet, and P. Gori-Giorgi.
{ \it Phys. Rev. B }
{\bf 2014 }, {\it 89}, 125106.

\end{thebibliography}
\end{document}